\documentclass[conference,10pt]{IEEEtran}
\IEEEoverridecommandlockouts

\usepackage[utf8]{inputenc}
\usepackage{microtype}
\usepackage[colorlinks=true,urlcolor=blue,citecolor=blue,linkcolor=blue]{hyperref}
\usepackage{cite}
\usepackage{braket}
\usepackage{algorithm}
\usepackage{algpseudocode}
\usepackage{amsmath,amssymb,amsfonts}
\usepackage{graphicx, xcolor}
\usepackage{textcomp}

\begin{document}

\title{A SAT approach to the initial mapping problem in SWAP gate insertion for commuting gates}

\author{\IEEEauthorblockN{Atsushi Matsuo}
\IEEEauthorblockA{\textit{IBM Quantum} \\
\textit{IBM Research -- Tokyo} \\
\textit{Ritsumeikan University}\\
Tokyo, Japan \\
matsuoa@jp.ibm.com}
\and
\IEEEauthorblockN{Shigeru Yamashita}
\IEEEauthorblockA{\textit{Ritsumeikan University} \\
Shiga, Japan \\
ger@cs.ritsumei.ac.jp}
\and
\IEEEauthorblockN{Daniel J. Egger}
\IEEEauthorblockA{\textit{IBM Quantum} \\
\textit{IBM Research -- Z\"urich}\\
R\"uschlikon, Switzerland \\
deg@zurich.ibm.com}
}

\maketitle

\begin{abstract}
Most quantum circuits require SWAP gate insertion to run on quantum hardware with limited qubit connectivity.
A promising SWAP gate insertion method for blocks of commuting two-qubit gates is a predetermined swap strategy which applies layers of SWAP gates simultaneously executable on the coupling map.
A good initial mapping for the swap strategy reduces the number of required swap gates.
However, even when a circuit consists of commuting gates, e.g., as in the Quantum Approximate Optimization Algorithm (QAOA) or trotterized simulations of Ising Hamiltonians, finding a good initial mapping is a hard problem.
We present a SAT-based approach to find good initial mappings for circuits with commuting gates transpiled to the hardware with swap strategies.
Our method achieves a 65\% reduction in gate count for random three-regular graphs with 500 nodes.
In addition, we present a heuristic approach that combines the SAT formulation with a clustering algorithm to reduce  large problems to a manageable size.
This approach reduces the number of swap layers by 25\% compared to both a trivial and random initial mapping for a random three-regular graph with 1000 nodes.
Good initial mappings will therefore enable the study of quantum algorithms, such as QAOA and Ising Hamiltonian simulation applied to sparse problems, on noisy quantum hardware with several hundreds of qubits. 
\end{abstract}


\section{Introduction}

Quantum computers offer a new paradigm to address complex problems by processing information with the laws of quantum mechanics~\cite{Shor1994, Moll2018}.
In practice, quantum computing architectures have a limited qubit connectivity which is expressed as a \emph{coupling map}, i.e., a graph in which nodes are physical qubits and edges are hardware native two-qubit gates.
For example, superconducting qubits are arranged in a two-dimensional lattice such as a grid~\cite{Harrigan2021} or a heavy-hex graph~\cite{Chamberland2020}.

Most quantum circuits require gates that are not hardware native and must be decomposed into native gates~\cite{Iwama2002}.
Furthermore, SWAP gates are inserted in these circuits to overcome the limited qubit connectivity~\cite{Siraichi2018}.
These tasks are done by a \emph{transpiler} which maps program circuits into hardware executable circuits.
The transpiler must therefore (i) select the physical qubits to work with (ii) perform an \emph{initial mapping} of program qubits to the selected physical qubits and (iii) decompose the program gates and insert SWAP gates such that all program gates are implemented by hardware native gates.
On noisy hardware the physical qubits can be selected by their properties such as coherence times, gate fidelities and connectivity~\cite{Tannu2019, Carrera2022}.
Adding SWAP gates introduces noise making it desirable to minimize their number.
This minimization, done in steps (ii) and (iii), is, however, a hard combinatorial optimization problem~\cite{Lye2015, Murali2019}.
Optimal solutions to this problem can only be obtained for small quantum circuits~\cite{Lye2015, Siraichi2018}. 
Heuristic strategies have therefore been developed~\cite{Kole2016, Bhattacharjee2018, Li2019, Tannu2019, ITOKO202043} and some included in transpilers such as Qiskit~\cite{Qiskit} and tKet~\cite{Sivarajah2020}.
Leveraging pulse-level information also helps transpilers reduce noise~\cite{Alexander2020, Earnest2021}.

The initial mapping problem is NP-complete and equivalent to a subgraph isomorphism~\cite{Maslov2008}.
A good initial mapping significantly reduces the gate count~\cite{Siraichi2018} and many heuristics have been developed to produce them. 
For example, by assigning program qubits to physical qubits based on connectivity properties~\cite{Maslov2008, Siraichi2018, Murali2019, Alam2020}.
For two-dimensional grid coupling maps a placement strategy can be devised using how often program qubits interact~\cite{Bhattacharjee2018}.
The SABRE (SWAP-based BidiREctional) heuristic algorithm uses circuit reversibility to iterate the initial mapping and reduce SWAP gate count~\cite{Li2019}.

Structured circuits made of blocks of commuting two-qubit gates, such as QAOA~\cite{Farhi2014} and trotterized Ising simulation~\cite{Carrera2022}, are prevalent in quantum computing applications.
Optimal SWAP gate insertion is still a hard problem even when exploiting commutativity~\cite{Lao2021, Alam2020}.
A promising method which performs well for dense problems is to use a predetermined \emph{swap strategy} which applies layers of SWAP gates simultaneously executable on the coupling map~\cite{Weidenfeller2022}.
To map a given circuit to the hardware with swap strategies the transpiler will, for a given initial mapping, loop through the swap layers and apply any two-qubit gate that is feasible on the current qubit configuration.
As the problem density increases the initial mapping becomes less relevant.
However, as we show in this work, a good initial mapping significantly reduces the number of swap layers for problems that are not fully connected.
Quantum computers are soon expected to have a few hundreds of noisy qubits~\cite{Roadmap}.
Testing, e.g., QAOA on such quantum computers will first be done with sparse graphs close to being hardware native due to the limited gate fidelity~\cite{Weidenfeller2022}.
A procedure to generate good initial mappings for a few hundreds of qubits and sparse problems is thus needed.

\begin{figure*}[t]
    \centering
    \includegraphics[width=\textwidth]{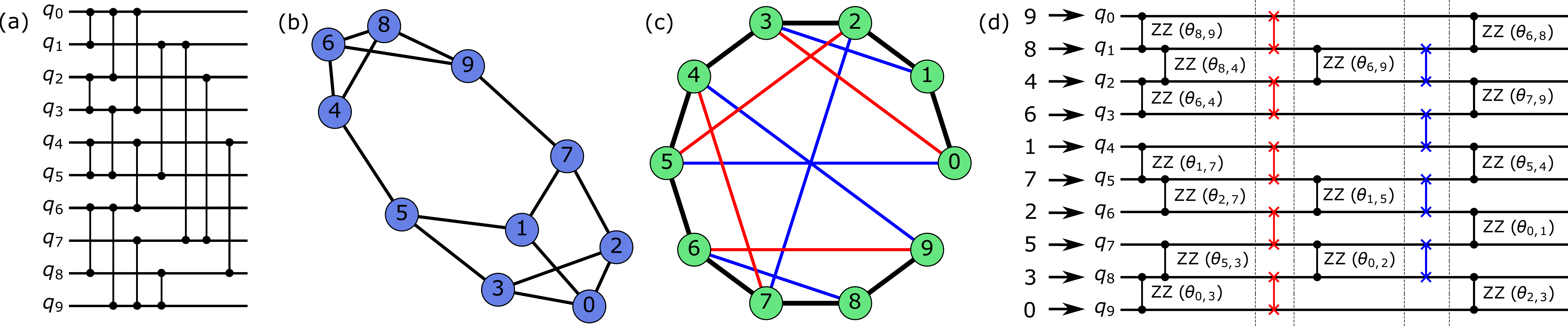}
    \caption{
    Illustration of an initial mapping.
    (a) A quantum circuit made of commuting two-qubit $ZZ$ gates and (b) its corresponding program graph.
    Each node is a program qubit and each edge a $ZZ$ gate.
    (c) Connectivity graph after two layers of a line swap strategy on ten linearly connected qubits.
    (d) Quantum circuit of the program graph in (a) and (b) with SWAP gates inserted to match the connectivity graph in (c).
    The arrows indicate the initial mapping of the program qubits to physical qubits.
    Edge $(i, j)$ in the program graph in (b) corresponds to a controlled phase gate with angle $\theta_{i,j}$ in (d). The layers of SWAP gates in
red and blue correspond to $s_1$ and $s_2$ in a line swap strategy, respectively.
    The vertical barriers serve as a guide to the eye.
    }
    \label{fig:example}
\end{figure*}

{\bf Our Contribution.} We propose a method to find the optimal number of swap layers even for sparse problems.
First, we observe that a good initial mapping of qubits, which is not considered in~\cite{Weidenfeller2022}, significantly reduces the gate count for circuits that are not fully connected. 
Then, we formulate this problem as a subgraph isomorphism.
However, we confirmed that the resulting subgraph isomorphism problem cannot be handled by an established solver like VF2~\cite{Cordella2004} in a practical time.
We thus formulate the subgraph isomorphism problem as a SAT problem to leverage powerful SAT solvers. 
Our method drastically reduces the gate count for a large problem with 500 qubits.
Furthermore, we propose a clustering-based heuristic for even larger problems that we demonstrate with 1000 qubits in a reasonable time. 

We give a brief introduction to quantum computing in Sec.~\ref{sec:qc_intro}.
In Sec.~\ref{sec:init_mapping} we describe the initial mapping problem for swap strategies which we formulate as a SAT problem in Sec.~\ref{sec:sat_mapping}.
We show a binary search algorithm capable of finding good initial mappings for problems with up to 500 qubits in Sec.~\ref{seq:sat_binary}.
In Sec.~\ref{sec:heuristic} we present the heuristic algorithm that decomposes large initial mapping problems to show a gate count reduction for a 1000 qubit problem.
We conclude in Sec.~\ref{sec:conclusion}.

\section{quantum computing\label{sec:qc_intro}}

A quantum circuit, as shown in Fig.~\ref{fig:example}(a), is a model of a quantum computation.
Each horizontal line is a qubit, i.e., a two-level quantum system with a state $\alpha\ket{0}+\beta\ket{1}$ such that $\alpha,\beta\in\mathbb{C}$ and $|\alpha|^2+|\beta|^2=1$ to conserve probability.
Physically, qubits are implemented by quantum systems with two controllable levels.
As example, a transmon qubit is engineered by a non-linear inductor in parallel with a capacitor~\cite{Koch2007}.
Each symbol between qubits is a quantum gate, i.e., a unitary matrix. 
Different quantum architectures implement different gate sets called the hardware native gates.
For example, fixed-frequency qubits coupled by microwave resonators implement the two-qubit CNOT gate with a cross-resonance interaction that drives a control qubit at the frequency of a target qubit~\cite{Sheldon2016}.
A universal set of gates, such as single-qubit rotations with the CNOT gate, can build any quantum gate. 
Due to engineering constraints two-qubit gates cannot be applied between arbitrary qubits.
This is described by a \emph{coupling map}, i.e., a graph where each node is a physical qubit and each edge is a hardware native two-qubit gate. 
A quantum circuit that requires gates between non-neighboring qubits in the coupling map is \emph{transpiled} to the hardware by inserting SWAP gates to swap program qubits.

\section{Mapping program qubits to physical qubits\label{sec:init_mapping}}

A swap strategy $\mathcal{S}=(S, \vec{o}, C_0)$ is a set $S=\{s_1, ..., s_m\}$ of $m$ different layers of simultaneously applicable SWAP gates on a hardware coupling map $C_0$ and an ordering $\vec{o}=(o_1, ..., o_L)$ in which to apply the swap layers. Each layer of simultaneously applicable SWAP gates $s_i \in S$ is a list of pairs of physical qubits on which the SWAP gates are applied. $o_j=i\in \{1,...,m\}$ implies that the $j^\text{th}$ layer of SWAP gates is $s_i\in S$.
As an example, a common line swap strategy is $(\{s_1, s_2\}, (1, 2, 1, 2,...), C_0)$, where $s_1 = ((0, 1), (2, 3), ...)$ and $s_2 = ((1, 2), (3, 4), ...)$. 
Here, we alternate swap layers on even and odd edges\footnote{We call an edge $(i,i+1)$ of a linear coupling map even if $i = 0 \mod 2$ and odd otherwise.} of $C_0$ and we achieve full connectivity in $n - 2$ swap layers which is provably optimal~\cite{Weidenfeller2022}.
This implies that in swap layer $s_1$ we apply SWAP gates to qubits 0 and 1, and qubits 2 and 3, and so on.
The ordering $\vec{o}$ is $(1, 2, 1, ...)$, meaning that we alternate between $s_1$ and $s_2$.
For a line of $n$ physically connected qubits, the coupling map $C_0$ is a list of pairs $(i, i+1)$ with $i=0,...,n-2$. Figure~\ref{fig:example}(d) exemplifies the first two layers of a line swap strategy on 10 qubits. Here, the layers of SWAP gates in red and blue correspond to $s_1$ and $s_2$, respectively.

After $l$ layers of the swap strategy, the hardware coupling map is transformed into the effective \emph{connectivity graph} $C_l$, see Fig.\ref{fig:example}(c). Here, the black edges indicate the original edges of the linear coupling map $C_0$.
The red and blue edges indicate new connections introduced by the first and second layers of SWAP gates, respectively, as shown in Fig.\ref{fig:example}(d) with matching colors. 
For instance, applying the first swap layer $s_1$ makes $q_0$ and $q_3$ adjacent since $s_1$ swaps qubits 0 and 1 as well as qubits 2 and 3. 
There is thus a red edge between qubits 0 and 3 in Fig.\ref{fig:example}(c) because the physical qubits 1 and 2 are connected in $C_0$. 
A swap strategy achieves full connectivity if $C_L$ is a complete graph.

We describe a block of commuting two-qubit gates by a graph $P=(V, E)$ which we call the \emph{program graph}.
Each edge corresponds to a two-qubit gate and each node is a program qubit which can be a decision variable in a QAOA.
For incomplete program graphs a carefully chosen initial mapping reduces the circuit depth for a given swap strategy.
An example of a random three-regular problem mapped and transpiled to a line coupling map is shown in Fig.~\ref{fig:example}.
Here, $\mathcal{S}$ is a line swap strategy.
A trivial mapping $i\to q_i$ requires a circuit with eight swap layers of $\mathcal{S}$ since the program qubits $0$ and $2$ move in the same direction in the line swap strategy. Crucially, an optimized initial mapping needs only two swap layers, see Fig.~\ref{fig:example}(d).

\begin{figure}[b]
    \centering
    \includegraphics[width=0.9\columnwidth]{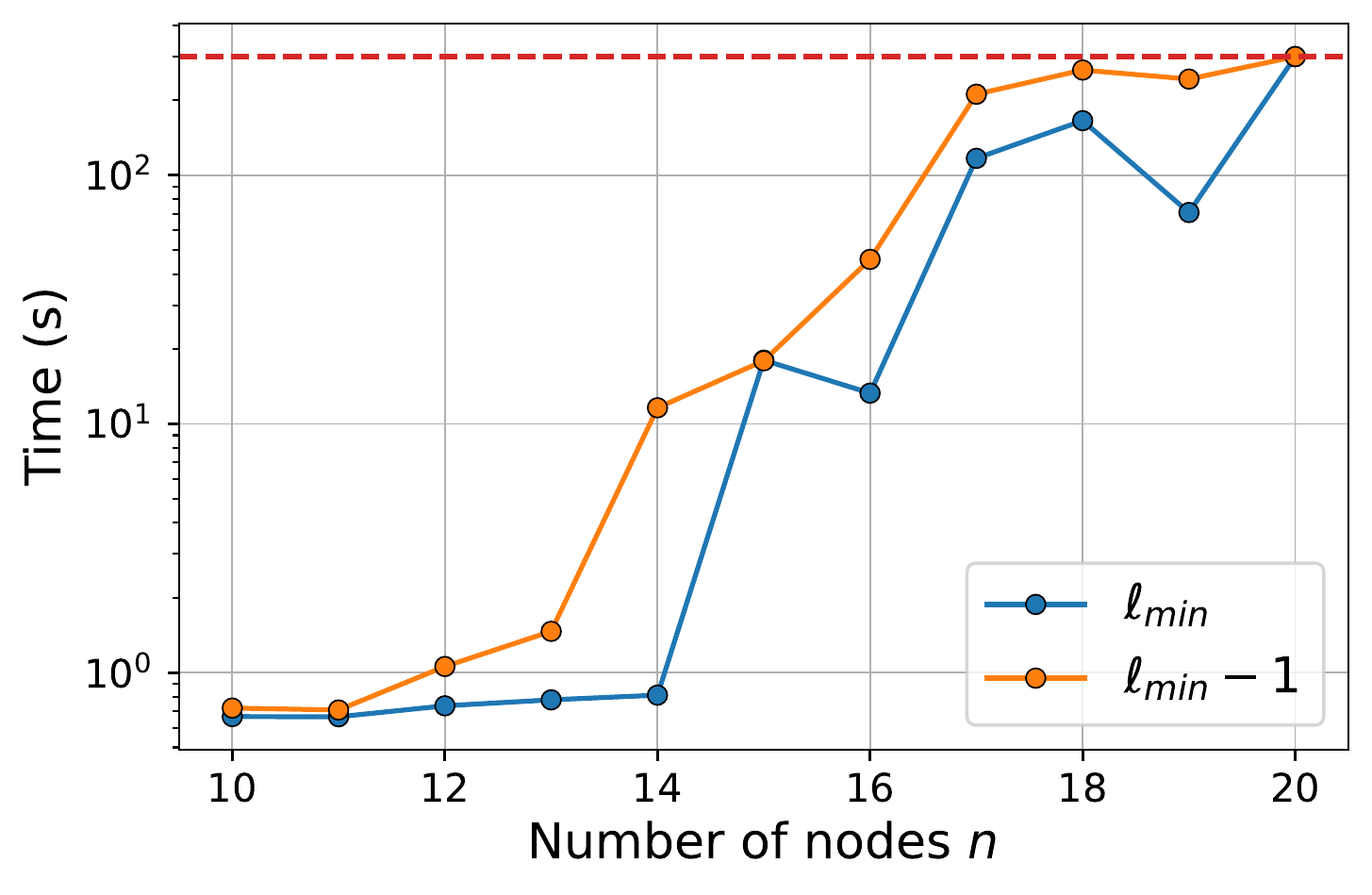}
    \caption{Time taken by the VF2 algorithm to map a $n$ node program graph $P_n$ to two connectivity graphs $C_{l_{min}}$ and $C_{l_{min} -1}$.
    The red horizontal line is the allotted time to the VF2 algorithm.}
    \label{fig:vf2}
\end{figure}

The initial mapping problem is a subgraph isomorphism problem where we wish to embed $P$ in $C_l$.
In this work we assume a given swap strategy $\mathcal{S}$ and seek the initial mapping $\pi:i\to q_i$ that assigns program qubit $i$ of $P$ to physical qubit $q_i$ in $C_0$ such that the least number of swap layers is needed. 
I.e., find the minimum $l$ such that $C_l$ can embed $\pi(P)$.

We numerically investigate\footnote{All the results presented in this work were obtained with a MacBook Pro with Apple M1 Max and 64GB memories.} the time taken by the VF2 solver to solve instances of the initial mapping problem with random program graphs with $n\in\{10,...,20\}$ nodes where each edge has a 20\% probability of occurrence.
For each $n$, five random graphs are mapped to the connectivity graph of a line swap strategy.
We run VF2 twice: once with the smallest number of swap layers capable of embedding each graph, labeled $l_\text{min}$, and once with $l_\text{min}-1$.
The time taken by VF2 grows quickly, see Fig.~\ref{fig:vf2}, making subgraph isomorphism solvers a poor choice to design good initial mappings for large problems. 
This is not surprising since the subgraph isomorphism problem is NP-complete in the general case and thus hard to solve.
In a conventional subgraph isomorphism problem we seek to embed a small graph in a comparatively larger graph.
However, in the initial mapping problem, the number of nodes in $P$ and $C_l$ are the same.
Furthermore, in our case, the connectivity graphs $C_0, C_1, \dots, C_{n-1}$ are fixed and there are known``easy" instances such as $C_{n-2}$ (complete graph) and $C_0$ (line graph) for a line swap strategy. Although we do not have theoretical proofs, our experimental results indicate that finding the smallest $k$ for $C_k$ requires exponential time. These graphs may not have a trivial shape, and therefore, finding the smallest $k$ may be still NP-hard in our case.

\section{A SAT formulation of the initial mapping\label{sec:sat_mapping}}

We now map the subgraph isomorphism problem in Sec.~\ref{sec:init_mapping} to a SAT problem~\cite{Toran2013} which allows us to benefit from advances in SAT solvers~\cite{Balyo2021, imms-sat18} which have also been leveraged to insert SWAP gates in quantum circuits~\cite{Hattori2018, matsuo2019}.
The SAT problem is expressed by literals $x_{i, j}$.
Here, $x_{i,j}$ is true if node $i$ in a program graph $P=(V,E)$ is mapped to node $j$ in a connectivity graph $C_l=(V_l,E_l)$ which is obtained after $l$ swap layers.
There are therefore $|V||V_l|$ literals, i.e., $n^2$ literals if $|V|=|V_l|=n$.
For example, when $x_{1,3}$ is true node 1 in the program graph is mapped to node 3 in the connectivity graph. 
Three conditions must be satisfied to embed the program graph into the connectivity graph $C_l$ such that all operations in a circuit can be performed in at most $l$ swap layers.

\textbf{Condition~1.} Each node $i$ in the program graph is assigned to only one node in the connectivity graph.
This condition is expressed as two sub-conditions. 
First, $i$ is assigned to at least one node which implies that at least one $x_{i,j}$ is true, i.e., the logical OR of all $x_{i,j}$ for $j=0,...,|V_l|-1$ is true
\begin{equation}
\label{eq:sat1}
x_{i,0} \lor x_{i,1} \cdots \cdots \lor x_{i,|V_l|-1} = 1.
\end{equation}
Second, $i$ is assigned to at most one node in the connectivity graph.
For example, we do not want $x_{i,j}$ and $x_{i,k}$ with $j\neq k$ to be simultaneously true, i.e.,
\begin{equation}
\label{eq:sat2}
\lnot x_{i,j} \lor \lnot x_{i,k}=1\quad \text{for}\quad j\neq k.
\end{equation}
Therefore, to assign node $i$ to at most one node in the connectivity graph the clause in Eq.~\eqref{eq:sat2} must be true for all pairs of nodes in $V_l$, i.e.,
\begin{eqnarray}
\label{eq:sat3}
\bigwedge_{j> k} \lnot x_{i,j} \lor \lnot x_{i,k}=1.
\end{eqnarray}
If  Eq.~\eqref{eq:sat1} and~\eqref{eq:sat3} hold for all $i$ then condition 1 is true.
Condition 1 thus generates $|V|$ clause of length $|V_l|$ and $|V||V_l|(|V_l|-1)/2$ clauses of length 2.

\textbf{Condition~2.} At most one node in a program graph is assigned to a node in a connectivity graph.  
For node $k\in V_l$ the condition $\lnot x_{i,k} \lor \lnot x_{j,k}=1$ prohibits simultaneously assigning nodes $i$ and $j$ to $k$.  
The conjunction of such clauses over all $i\neq j$ prevents assigning multiple nodes of $P$ to a node $k$ in $C_l$
\begin{eqnarray}
\label{eq:sat5}
\bigwedge_{i> j} \lnot x_{i,k} \lor \lnot x_{j,k}=1.
\end{eqnarray}
Condition 2 thus creates $|V_l||V|(|V|-1)/2$ clauses of length 2.

\textbf{Condition~3.} Adjacent nodes in a program graph must be adjacent after they are mapped to nodes in the connectivity graph.
Accordingly, for an edge $(i,j)\in E$ then $x_{i,k}$ being true implies that there must be a $x_{j,k'}$ with $(k,k') \in E_l$ that is also true. 
This implication is expressed as
\begin{equation}
\label{eq:sat6}
\lnot x_{i,k}\lor\left(\bigvee_{(k,k') \in E_l} x_{j,k'}\right).
\end{equation}
We can thus express condition 3 by taking the conjunction of all clauses (\ref{eq:sat6}) generated by each edge $(i,j) \in E$.
Condition 3 thus generates $|E||V_l|$ clauses of variable length.

The conjunction of the clauses of Conditions~1, 2 and 3 yields a SAT formulation of the initial mapping problem.
A SAT solver can therefore determine if there exists an initial qubit placement such that all gates in the circuit can be implemented by at most $l$ swap layers of the swap strategy.
If such a placement exists the SAT solver also returns a satisfying variable assignment.

\section{Finding the best initial mapping based on the SAT formulation\label{seq:sat_binary}}

We solve the initial mapping problem with PySAT~\cite{imms-sat18}, a Python library designed to solve SAT problems. 
PySAT provides built-in SAT solvers with low-level implementations such as C++, making it fast and easy to use.
In the experiments, we use a linear coupling map and the line swap strategy described in Sec.~\ref{sec:init_mapping} to represent the limited qubit connectivity in current quantum devices.

\begin{figure}[tb]
    \centering
    \includegraphics[width=\columnwidth]{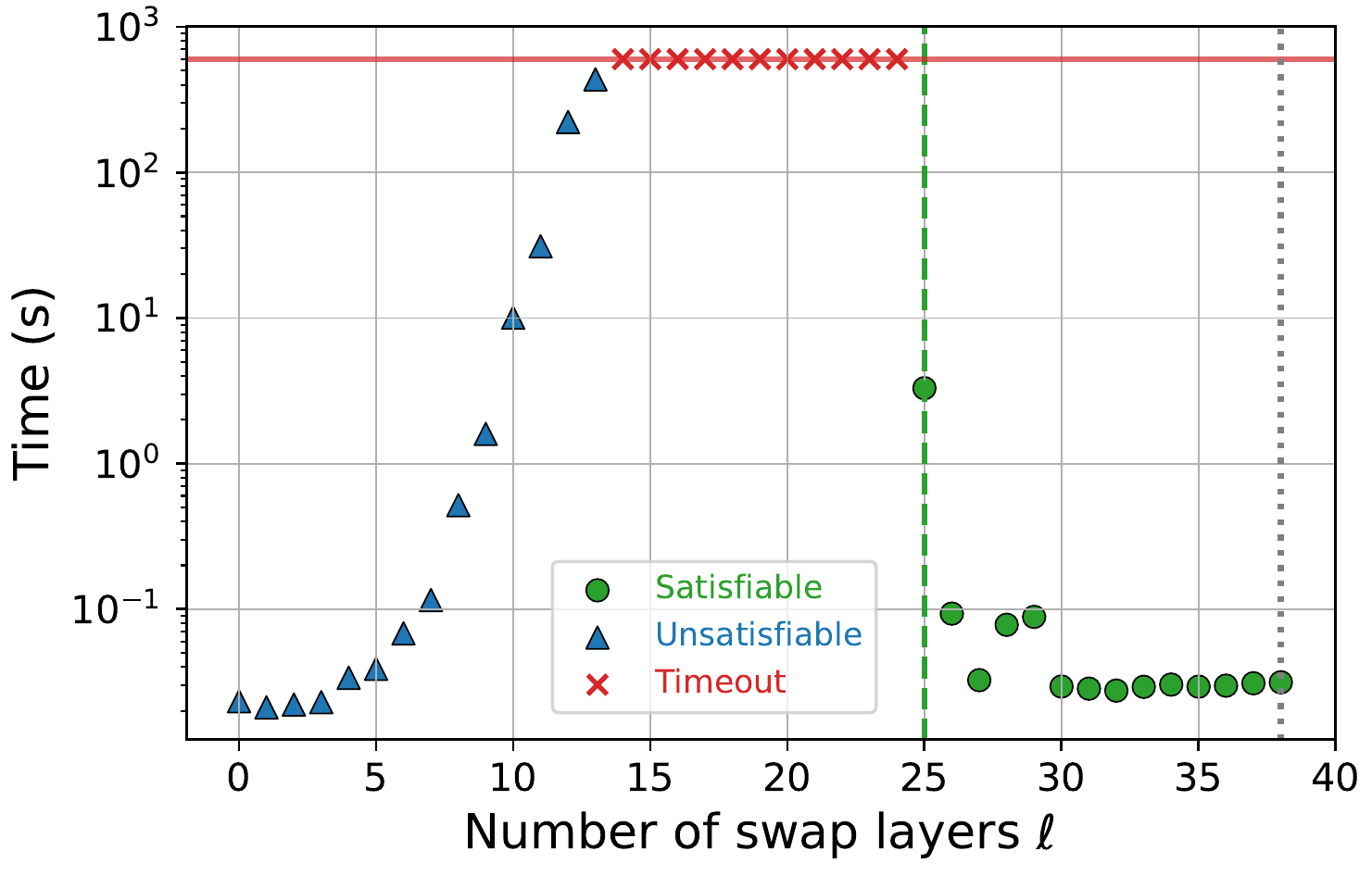}
    \caption{Time taken by PySAT to determine if  a 40 node graph can be embeded in a connectivity graph as a function of the number of swap layers.
    The red horizontal line indicates the allotted time to the SAT solver.
    The green dashed vertical line shows the resulting $l$.
    The grey dotted line shows the number of swap layers required by a trivial mapping.}
    \label{fig:figure2}
\end{figure}

As a preliminary experiment, we solve the initial mapping problem for a random graph with 40 nodes,
where each edge has a 20\% probability of occurrence.
We chose a sparse graph since swap strategies without an initial mapping do not perform well on them~\cite{Weidenfeller2022}.
We allow PySAT 600 seconds\footnote{We chose 600 seconds to make the optimization time manageable on a laptop. In a high-performance computing environment more resources may be allotted to the SAT solver.} to determine if an instance is satisfiable.
Since we transpile the program graph to a linear coupling map we solve 39 SAT instances; one for each $l\in\{0,...,38\}$. 
The SAT instances with a number of swap layers $l<14$ are not satisfiable, see Fig.~\ref{fig:figure2}, i.e., more swap layers are needed to overcome the limited qubit connectivity.
The SAT instances with $l>25$ are satisfiable and the circuit can be executed on the hardware.
PySAT cannot determine if the problem is satisfiable in 10 minutes for $l\in\{14, ...,24\}$.
We therefore observe the typical easy-hard-easy pattern of SAT instances~\cite{Gent1994, McCreesh2018}.
Crucially, the satisfiable SAT instance at $l=25$ reduces the number of swap layers by 34\% compared to a trivial initial mapping.
Note that the specific numbers $14$ and $25$ depend on the graph instance, but the easy-hard-easy pattern does not.

Based on these observations we can find a good initial mapping to reduce the number of swap layers with a binary search over $l$.
Since the number of swap layers grows linearly with problem size~\cite{Weidenfeller2022} we need only solve $\mathcal{O}(\log n)$ SAT instances.
The initial points of the binary search are $l_L=0$ and $l_R=|V|-2$, which are typically satisfiable and unsatisfiable, respectively, and easy to solve, e.g., see Fig.~\ref{fig:figure2}.
Furthermore, we allow the SAT solver a fixed time to determine if a SAT instance is satisfiable.
If the SAT solver cannot find a solution in the allowed time we consider the SAT instance to be unsatisfiable.
If $\mathcal{S}$ can reach full connectivity we are guaranteed to find an initial mapping in the worst case with an $l$ equivalent to full connectivity. 
Once the binary search has converged we make an initial mapping based on the solution of the last satisfiable SAT instance.
The algorithm is summarized below.

\begin{algorithm}[H]
\caption{The algorithm to find a good initial qubit mapping.}
\label{alg:init_layout_sat}
\begin{algorithmic}[1]
\Require program graph $P=(V,E)$, swap strategy $\mathcal{S}=(S, \vec{o}, C_0)$
\Ensure Initial layout that reduces the number of swap layers of $\mathcal{S}$ needed to execute $P$ on the hardware.
\State $l_L \gets 0$
\State $l_R \gets |V|-2$
\While{$l_L < l_R$}
    \State $l \gets \lfloor (l_L+l_R)/2 \rfloor$
    \State Create connectivity graph $C_l$ from $\mathcal{S}$
    \State Create SAT instance $\text{SAT}_\text{embed} (P, C_l)$ 
    \If{$\text{SAT}_\text{embed} (P, C_l)$ is satisfiable}
        \State $l_R \gets l$
    \Else
        \State $l_L \gets l+1$
    \EndIf
\EndWhile
\end{algorithmic}
\end{algorithm}

We test Algorithm~\ref{alg:init_layout_sat} on two types of sparse random graphs commonly used as benchmarks \cite{Weidenfeller2022, PhysRevLett.125.260505, PhysRevX.10.021067}: random three-regular graphs ($\text{RR3}_n$) and random graphs in which edges appear with a $20\%$ probability ($\text{Rand}_n$). We conduct experiments on graphs with $n=40$, $100$, $200$, and $500$ nodes and linear coupling maps. 
It is worth noting that Algorithm~\ref{alg:init_layout_sat} works with any coupling map and swap strategy.
For each graph size $n$, we perform experiments on five different graph instances.
We compare Algorithm~\ref{alg:init_layout_sat} to a random initial mapping and a trivial mapping, which maps program qubit $v_i$ in a program graph to physical qubit $v'_i$ in a connectivity graph.
We average over 100 different random initial mappings for each graph and chose the result with the minimum number of swap layers.

\begin{figure}[t]
    \centering
    \includegraphics[width=1.0\columnwidth]{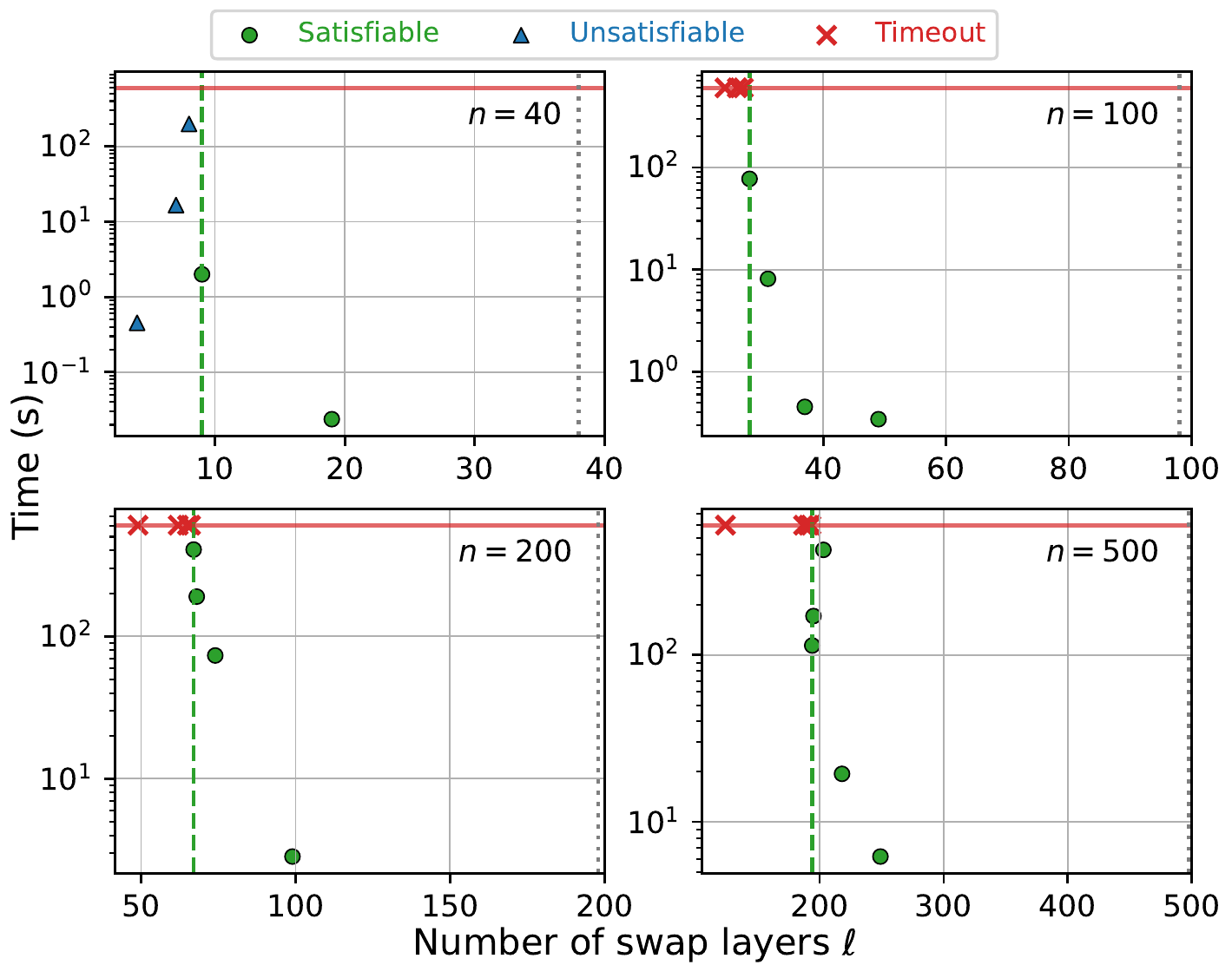}
    \caption{Time taken of each iteration of the binary search to map three-regular graphs with a different number of nodes $n$.
    The red horizontal line indicates the allotted time to the SAT solver.
    The green dashed vertical line shows the resulting $l_{min}$.
    The grey dotted line shows the number of swap layers required by a trivial mapping.
    }
    \label{fig:figure3}
\end{figure}

We allow PySAT a maximum of $600$ seconds per SAT instance $\text{SAT}_\text{embed}(P,C_l)$.
We observe a significant reduction in the number of SWAP layers needed, see Fig.~\ref{fig:figure3}.
The optimal $l$ may exist in the SAT instances which timed-out.
Crucially, it is possible to find a good yet suboptimal $l$ that significantly decreases the number of swap layers, and the number of CNOT gates in a practical time.
For example, Algorithm~\ref{alg:init_layout_sat} identifies an initial mapping for a 500 node three-regular graph that reduces the number of swap layers to less than 200 in 90 minutes while a trivial mapping requires full connectivity.
The SAT approach significantly decreases the number of CNOT gates and circuit depth as shown in Table~\ref{tab:sat_mapping_summary} which lists the number of swap layers, and the number of the CNOT gates averaged over the five graph instances at each size.
The trivial mapping nearly always requires full connectivity, i.e., $l=n-2$ swap layers.
Random initial mappings only make minor improvements although the best result was chosen from 100 different trials.
We do not expect this situation to change with a polynomial increase in the number of trials since the search space scales combinatorially with $n$.

We also apply the SABRE layout, implemented in Qiskit~\cite{Qiskit}, to find an initial mapping before inserting swap gates with a swap strategy.
SABRE layout finds an initial mapping with an iterative bidirectional routing of the program circuit~\cite{Li2019} for general swap insertion methods.
Since the SABRE layout method is not tailored to the swap strategy its results are as bad as the trivial mapping and the random mapping, see Table~\ref{tab:sat_mapping_summary}.
This result emphasizes the need for an initial mapping method tailored to the swap strategy.
We also compare the SWAP gate depth of a swap strategy with an initial mapping to SABRE swap routing~\cite{Li2019}, a heuristic swap insertion algorithm, applied after the SABRE layout.
The numbers of swap layers of Sabre swap routing, counted as the SWAP gate depth of the resulting quantum circuit, is much larger than the swap strategy since SABRE swap routing does not fully utilize the $ZZ$ gate commutativity.
We note that each layer of SWAP gates inserted by SABRE swap routing is less dense than in a swap strategy.
However, circuit depth is crucial for current noisy devices since their coherence time is limited.

The reduction from the SAT approach is especially large for random three-regular graphs since they are sparser than random graphs.
These results highlight that a good initial mapping significantly reduces the number of swap layers.
Furthermore, since the number of SWAP gates in each layer of a swap strategy linearly increases with the number of qubits $n$, decreasing the number of swap layers decrease the total number of CNOT gates more in the larger program graphs.

\begin{table*}[t]
    \caption{Number of swap layers and CNOT gates needed to execute a program graph on the hardware following different initial mappings obtained with the trivial, random, SABRE layout, and SAT strategies. The ``Sabre layout'' column is the number of layers of the line swap strategy after an initial mapping found by SABRE layout.
    The ``SABRE swap'' column shows the swap depth of a circuit after SABRE layout and SABRE swap routing.
    Each swap layer has, up to edge effects, $n/2$ SWAP gates and each SWAP gate requires three CNOT gates.
    $\eta$ is the ratio of the number of CNOT gates found with the SAT solver to the random initial mapping.
    }
    \label{tab:sat_mapping_summary}
    \centering
    \begin{tabular}{l | r r r r r | r r r r}\hline\hline
              & \multicolumn{5}{c}{Number of swap layers} & \multicolumn{4}{| c}{Number of CNOT gates} \\
              &         &          &        SABRE & SABRE & & & & \\
        Graph & Trivial & Random & layout & swap & SAT & Trivial & Random & SAT & $\eta$ \\ \hline
        $\text{RR3}_{40}$ & $38 \pm 0$ & $35\pm 1$ & $38 \pm 1$ & $52 \pm 12$ & $9 \pm 0$ & $2212 \pm 23$ & $2060\pm 44$ & $516 \pm 24$ & $0.25$\\
        $\text{Rand}_{40}$ & $38 \pm 0$ & $38 \pm 0 $& $38 \pm 0$ & $181 \pm 14$ &$25 \pm 1$ & $2223 \pm 0$ & $2200 \pm 28 $& $1454 \pm 79$ & $0.66$\\ \hline
        $\text{RR3}_{100}$ & $98 \pm 1$  & $95 \pm 1$ & $97 \pm 1$ &$224 \pm 54$ & $27 \pm 1$ & $14494 \pm 119$  & $14049 \pm 201$ & $4040 \pm 174$ & $0.29$\\
        $\text{Rand}_{100}$ & $98 \pm 0$ & $98 \pm 0$ & $98 \pm 0$ & $921 \pm 145$ & $84 \pm 1$ & $14553 \pm 0$ & $14553 \pm 0$ & $12475 \pm 94$ & $0.85$\\ \hline
        $\text{RR3}_{200}$ & $198 \pm 0$ & $195 \pm 0$ & $198 \pm 0$ & $680 \pm 66 $ &$68 \pm 2$ & $59103 \pm 0$ & $58149 \pm 120$ & $20239 \pm 513$ & $0.35$\\
        $\text{Rand}_{200}$ & $198 \pm 0$ & $198 \pm 0$ & $198 \pm 0 $& $ 2322 \pm 313 $&$183 \pm 0$ & $59103 \pm 0$ & $59103 \pm 0$ & $54746 \pm 146$ & $0.93$\\ \hline
        $\text{RR3}_{500}$ & $498 \pm 0$ & $495 \pm 1$ & $498 \pm 0$ & $4373 \pm 285$ & $196 \pm 6$ & $372454 \pm 366$ & $370358 \pm 872$ & $147006 \pm 4405$ & $0.35$\\
        $\text{Rand}_{500}$ & $498 \pm 0$ & $498 \pm 0$ & $498 \pm 0$ &$18290 \pm 14119$ &$-$ & $372753 \pm 0$ & $372753 \pm 0$ & $-$ & $-$\\ \hline\hline        
    \end{tabular}
\end{table*}

\section{Additional strategies for scalability\label{sec:heuristic}}

Algorithm~\ref{alg:init_layout_sat} presented in Sec.~\ref{seq:sat_binary} constructs a SAT problem with $n^2$ variables.
Furthermore, we could not run the 500 nodes random graph instances on the MacBook as they required more than 64~GB of memory and a single iteration of the binary search took more than 20 minutes.
To make the problem more manageable we introduce a heuristic to perform the initial mapping that iterateively maps sub-graphs of the program graph to the connectivity graph.
More formally, for a program graph $P=(V,E)$ with $n$ variables and a corresponding swap strategy $\mathcal{S}$ we select a sub-graph $P_0=(V_0, E_0)\subset P$ such that $|V_0|=n_0$.
We then find the $l_0$ with Algorithm~\ref{alg:init_layout_sat} that allows us to embed $P_0$ in $C_{l_0}$.
The resulting SAT problem has $n_0n$ decision variables with $n_0<n$.
We iterate the mapping.
At iteration $i$ we select a subgraph $P_i=(V_i, E_i)\subset P$ such that $P_i\cap (\cup_{j=0}^{i-1}P_j)=$ \O~and build the SAT problem such that the nodes of $P_i$ are mapped to unassigned qubits with the condition that any edge connecting a node from $V_i$ to $\cup_{j=0}^{i-1}V_j$ also has an edge in the connectivity graph $C_{l_i}$ of $\mathcal{S}$.
To find $l_i$ we perform a binary search in the interval $\{l_{i-1},...,L\}$ with $L=n-2$ for a line swap strategy.

\begin{figure}[tb]
    \centering
    \includegraphics[width=\columnwidth]{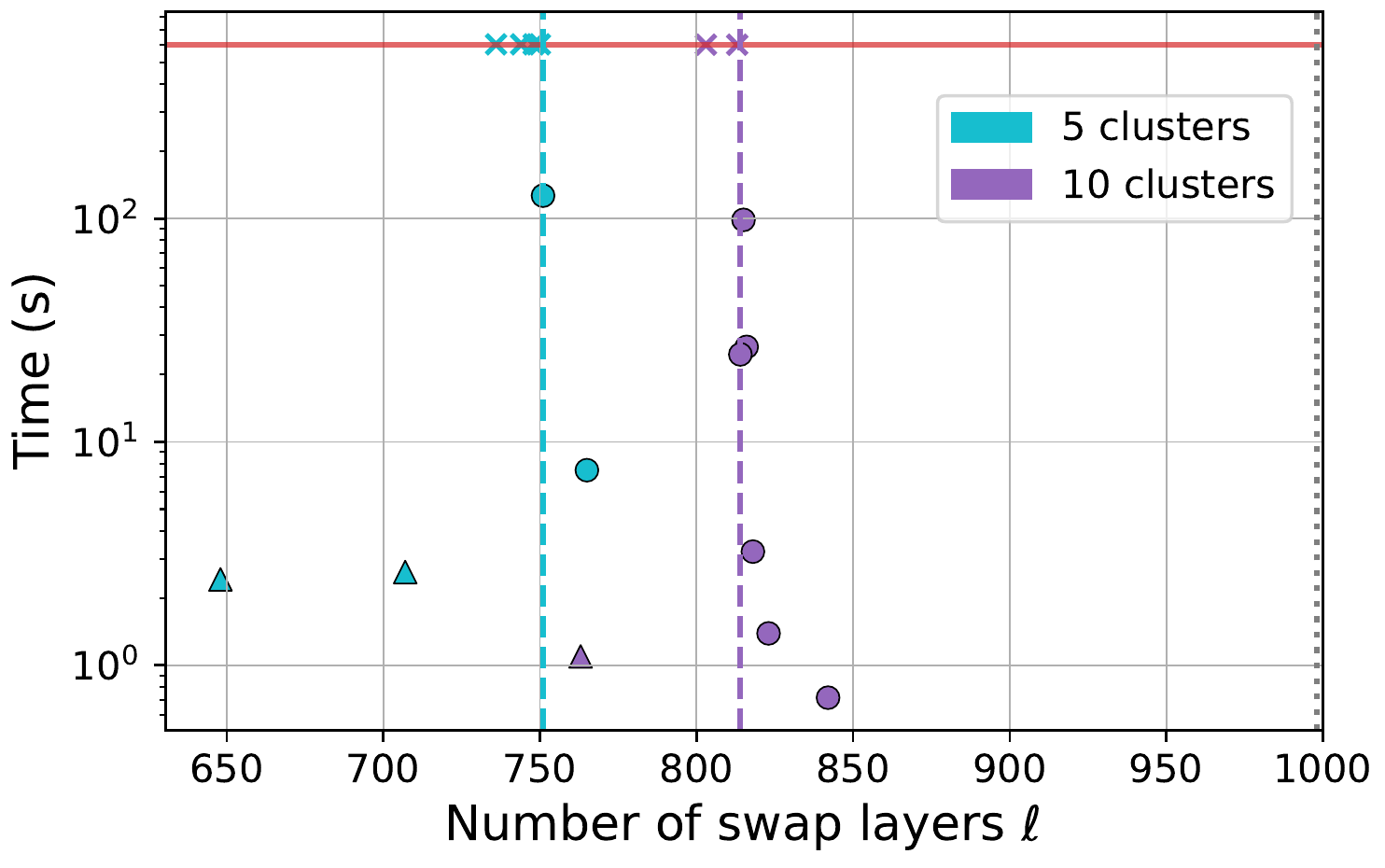}
    \caption{Time taken of each iteration of the binary search to map a three-regular graph with 1000 nodes to a linear coupling map. 
    Blue and purple markers correspond to data in which the sub-graphs $P_i$ have 200 and 100 nodes, respectively. 
    The circle, triangle, and cross markers correspond to satisfiable, unsatisfiable, and timeout, respectively.
    The red horizontal line indicates the allotted time to the SAT solver.
    The blue and purple dashed vertical lines show the resulting $l_{min}$ for the 200 and 100 node clustering, respectively.
    The grey dotted line shows the number of swap layers required by a trivial mapping.}
    \label{fig:1000_nodes}
\end{figure}

We test this heuristic on a random three-regular graph with 1000 nodes.
The sub-graphs $P_i$ are found with the spectral clustering~\cite{Jianbo2000} implemented in Scikit-learn~\cite{scikit-learn}.
We test the heuristic twice: once with ten clusters of 100 nodes and once with five clusters of 200 nodes.
Under these conditions the problem is manageable on a MacBook Pro and we identify an initial mapping that reduces the number of swap layers, see Fig.~\ref{fig:1000_nodes}.
With sub-graphs $P_i$ with 200 and 100 nodes we find a $l_\text{min}=751$ and $l_\text{min}=814$, respectively.
Smaller sub-graphs are computationally easier to manage but produce initial mappings with a larger number of swap layers.
This is expected since more clusters simplify the problem.

\section{Conclusion\label{sec:conclusion}}

Swap strategies can efficiently transpile circuits made of blocks of commuting two-qubit gates to hardware resulting in low-depth circuits~\cite{Weidenfeller2022}.
However, a good initial mapping further reduces the number of required swap gates for program graphs that are not complete.
When formulated as a subgraph isomorphism the initial mapping problem of embedding a program graph $P$ in a connectivity graph $C_l$ can only be solved for small instances with less than $\sim20$ nodes.
We therefore developed a SAT-based approach that finds good initial mappings for circuits with commuting gates that are transpiled to the hardware with swap strategies. 
We formulated the subgraph isomorphism as a SAT instance to benefit from the progress in SAT solvers.
A binary search reduces the number of swap layers by finding an initial mapping in $\mathcal{O}(\log n)$ steps.
We also proposed a heuristic approach to map graphs that are too large to be handled as a single SAT instance.
The heuristic approach divides the program graph into several clusters, and iteratively applies the binary search over the resulting smaller SAT instances. 
Our results show a significant decrease in the number of swap layers and CNOT gates for program graphs with up to 1000 qubits.

Future work may devise more efficient heuristics to solve the initial mapping problem on large instances.
Nevertheless, the methodology proposed here will allow us to map, e.g., sparse QAOA circuits to the quantum hardware to be developed in the coming years~\cite{Roadmap}.
The initial mapping will be crucial since the gate fidelity limits the structure of the program graph to sparse graphs that reassembles the coupling map~\cite{Franca2020, Weidenfeller2022}.

\bibliographystyle{IEEEtran}
\bibliography{main}

\end{document}